\documentstyle[12pt,amssymb]{article}    
\begin{document}

\title{Phase ordering at the lambda transition in liquid $^4$He}

\author{Torsten Fliessbach\thanks{Electronic address:
fliessbach@physik.uni-siegen.de} 
\\Fachbereich Physik, University of Siegen\\
 57068 Siegen, Germany}

\date{}

\maketitle

\begin{abstract}

A modified ideal Bose gas model is proposed as an approach for
liquid helium at the lambda point. The decisive modification of the
ideal Bose gas model is the use of phase ordered single particle
functions. The entropy due to this phase ordering is calculated. Its
statistical expectation value yields a logarithmic singularity of the
specific heat.

\vspace*{2cm}


\end{abstract}

\newpage

\section{Introduction}
\label{s1}
As suggested by London \cite{lo54} and substantiated by Feyn\-man
\cite{fe53} there is an intimate relation between the lambda transition
in $^4$He and the Bose-Einstein condensation of the ideal Bose gas
(IBG). The IBG explains \cite{pu74} some basic properties of liquid
$^4$He like for example the irrotational superfluid flow. We propose a
modification of the IBG that leads to a realistic expression for the
specific heat.

The single particle functions of the IBG are of the form
\begin{equation}
\label{e1}
\varphi_{\bf k} \propto
 \exp (\,{\rm i}\,{\bf k}\cdot{\bf r} ) =
 \exp [\,{\rm i}\,(k_1 x + k_2 y + k_3 z )] .
\end{equation}
Without changing the kinetic energy we may alternatively consider the
single particle functions
\begin{equation}
\label{e2}
\varphi^{\rm p.o.}_{\bf k} \propto
 \exp\big[\,{\rm i}\,(k_2 y + k_3 z)\big]\, 
\sin(q x + \phi) ,
\end{equation}
where $q=k_1$. For an arbitrary direction (chosen here as the $x$
direction) the functions $\varphi^{\rm p.o.}_{\bf k}$ have a specific
phase $\phi$; they are {\em phase ordered}\/ (p.o.). This phase
ordering implies the following correlation: For $k_1=k_1'=q$, the
probability densities $|\varphi^{\rm p.o.}_{\bf k}|^2$ and
$|\varphi^{\rm p.o.}_{{\bf k}'}|^2$ have common minima and maxima.
That means that two particles with the same momentum in one
extinguished direction have an additional positive spatial correlation.
No such correlations are present for the single particle functions
(\ref{e1}).

The phase ordering leads to a correlation energy of the form
\begin{equation}
\label{e3}
E_{\rm corr} \propto \sum_{q} \nu_q^{\, 2} 
\quad \mbox{with} \quad \nu_q = \sum_{k_2,\, k_3} n_{\bf k} 
.
\end{equation}
Here $n_{\bf k}$ is the number of atoms with momentum ${\bf k} =
(q,k_2,k_3)$ and $\nu_q$ is the number of atoms with common momentum
$q$ in $x$ direction. In the group of $\nu_q$ atoms each one has an
extra correlation with the other atoms; this results in a $\nu_q^{\,2}$
contribution to the energy. This correlation energy has been
investigated in detail in Ref.\  \cite{fl91}.

We will use the IGB many-body wave functions together with Jastrow
factors as proposed by Chester \cite{ch55}. The effective interaction on
the model level is then {\em attractive}\/ (the hard core is cut out by
the Jastrow factors). This means that the energy $E_{\rm corr}$ is
negative and that the correlations are favoured by the condition of
minimal free energy.

For a {\em finite}\/ correlation energy one must assume that the phase
ordering is {\em local}\/. The phase correlations of the single
particle functions (\ref{e2}) could be enforced by physical boundary
conditions for the considered macroscopic volume $V$ (instead of
periodic boundary conditions for the single particle functions
(\ref{e1})). In this case the correlations are a surface effect
implying $E_{\rm corr}\propto V^{2/3}$, and $E_{\rm corr}/V\to 0$ for
$V\to \infty$.  Assuming {\em local}\/ phase ordering introduces a
parameter $V_0$ that defines the finite range over which the single
particle functions are correlated. The directions ($x$ direction in
Eq.\ (\ref{e2})) of phase ordering are in general different
in various subvolumes. The statistical average over these subvolumes
ensures the isotropy of the bulk liquid. Evaluating the energy
(\ref{e3}) with the IBG expectation values $\langle n_{\bf k}\rangle$
yields a logarithmic singularity \cite{fl91}, i.e., $\langle E_{\rm
corr} \rangle \propto -t \ln |t|$, where $t$ is the relative
temperature.  Adjusting the amplitude of this singularity to the
experimental value fixes the model parameter $V_0$ (leading to $V_0/v
\approx 10^5 $, where $v$ is the volume per atom).

It is the aim of this paper to calculate the entropy change due to the
assumed local phase ordering. This calculation is independent of the
previous determination of the correlation energy (\ref{e3}). In order
to calculate the phase ordering entropy we establish a relation between
the functions (\ref{e2}) and coherent states; this relation is
analogous to Anderson's connection \cite{an66} between the condensate
wave function and locally coherent states. For coherent states the
phase variances are given by $\Delta \phi_{\bf k} =1/ (2\sqrt{n_{\bf
k}})$. Restricting the phases by $\Delta \phi_{\bf k}$ yields the phase
ordering entropy $S_{\rm p.o}(n_{\bf k})$. Using the occupation numbers
$\langle n_{\bf k}\rangle$ of the IBG form we find logarithmic
singularity for the thermodynamic entropy, $\langle S_{\rm p.o} \rangle
\propto -t\,\ln |t|$.  For the specific heat per particle, $c = -A \ln
|t|$, we obtain the result $A=3\,k_{\rm B}/2$, where $k_{\rm B}$ is
Boltzmann's constant.

As a theory of the lambda transition, the presented approach is a
modest attempt only. Instead of determining exact many-body states we
make a guess of the relevant correlations. Instead of evaluating the
partition sum we use the IBG expectation values (arguing that the
correlation effects are of minor influence on the occupation of the
individual levels). The determination of the exact many-body states
{\em and}\/ the exact evaluation of the partition sum is, of course,
not feasible. One might, however, try to find a suitable variational
ansatz for the many-body states for which the free energy can be
calculated and minimized (like it may be done for BCS states).  The
present evaluation of the entropy together with the previous evaluation
of the correlation energy (\ref{e3}) are a first step towards such an
approach. As it stands, we present a phenomenological many-body model
that yields a promising result, namely a logarithmic singularity with a
sensible strength.

As a many-body approach for the lambda transition the presented model
lies outside the main stream of realistic models for liquid helium.
There are several well-known and thoroughly studied microscopic
many-body approaches (for an overview see Ref.\  \cite{na98}),
like diagrammatic (Green's function) approaches \cite{ja85}, variational
methods (in particular the correlated basis function method and its
generalizations \cite{se92,ri95}), and Monte Carlo calculations. As far
as the approaches are based on elementary excitations they are
restricted to low temperatures. For temperatures around the lambda
point, there are successful applications of path-integral Monte Carlo
simulations \cite{po87,ce95} and of the variational approach \cite{ri99}.
These approaches do, however, not yield analytic and realistic
expressions for the asymptotic behavior. What is nowadays widely
considered as {\em the}\/ theory of the lambda transition and its
asymptotic properties is the renormalization group theory \cite{do87}.
In contrast to the model presented here, this theory is not a many-body
description.

Several authors \cite{la69,jo70,ka74,le82} have considered coherent
states in the description of liquid helium, too. Their investigations
concern mostly the connection between the condensate and coherent
states as in Anderson's paper \cite{an66}. The local phase ordering
that we consider is not specifically related to any of these other
investigations.

\section{Phase ordering entropy}
\label{s2}
\subsection{Many-body states}
\label{s2.1}
Chester \cite{ch55} studied the density matrix for many-body wave
functions of the form $\Psi = {\cal S}  F\Psi_{\rm IBG}$. Here $S$
denotes the symmetrization operator, $F = \prod_{i,j} f(r_{ij})$ is the
Jastrow factor, $\Psi_{\rm IBG}({\bf r}_j, n_{\bf k})$ describes the
IBG state, and ${\bf r}_j$ defines the positions of the helium atoms.
Using hard sphere factors for $f(r)$, Chester obtained the IBG
values $\langle n_{\bf k} \rangle_{\rm IBG}$ for the average occupation
numbers $\langle n_{\bf k}\rangle$.  McMillan \cite{mc65} used the
ansatz $\Psi = {\cal S}  F$ for the ground state; for realistic
atom-atom interactions he determined the function $f(r)$ by minimizing
the energy. This yields a realistic pair correlation function and a
realistic binding energy.

We follow basically Chester's approach replacing, however, the single
particle functions (\ref{e1}) by the phase ordered ones (\ref{e2}). The
many-body wave function is then of the form
\begin{equation}
\label{e4}
\Psi = {\cal S} F 
\prod_{j=1}^N \varphi^{\rm p.o.}_{{\bf k}_j} ({\bf r}_j)
= {\cal S} F 
\prod_{ {\bf k} } \left[\, \varphi^{\rm p.o.}_{\bf k} \,\right]^{n_{\bf
k}}  .
\end{equation}
The factor $[\varphi_{\bf k} ]^{n_k}$ in the last expression stands for
the product $\varphi_{\bf k}({\bf r}_{\nu+1}) \cdot \varphi_{\bf
k}({\bf r}_{\nu + 2})\cdot \ldots\cdot \varphi_{\bf k}({\bf
r}_{\nu+n_k})$. The total number of atoms is denoted by $N$.

As explained in the Introduction, a finite correlation energy
(\ref{e3}) requires finite extensions of the phase ordered single
particle functions. For this purpose the volume $V$ is thought to be
divided into $V/V_0$ subvolumes of size $V_0$, and the single particle
functions are restricted to such subvolumes. The phase ordering can 
then be obtained by requiring physical boundary conditions for the
subvolumes, or by considering coherent states in these subvolumes
(Sec.\ \ref{s2.3}).  The product over the subvolumes might be
explicitly displayed in the last expression of Eq.\ (\ref{e4}). We will
adjust the prefactor of the momentum integrals such that the result
corresponds to the total volume $V$.

The Jastrow factors are of no influence on the statistical counting.
Without phase ordering (i.e., using the single particle functions
(\ref{e1}) in Eq.\ (\ref{e4})) one obtains, therefore, the well-known
IBG expression for the entropy:
\begin{equation}
\label{e5}
S_{\rm IGB} = k_{\rm B} \sum_{{\bf k}} \left[ 
(1 +  n_{\bf k})\,\ln (1 +  n_{\bf k} )
 -  n_{\bf k} \,\ln (n_{\bf k} )\right]
.
\end{equation}
The entropy change due to the phase ordering may be divided into two
parts: 
\begin{enumerate}
\item[(a)]
All $\nu_q$ atoms in states with $k_1=q$ must adopt the same (mean)
phase. This leads to a replacement of the entropy (\ref{e5}) by a
modified expression $S'_{\rm IBG}$.
\item[(b)]
A well-defined phase requires a small phase variance. The entropy
change due to this restriction is denoted by $S_{\rm p.o.}$. This
decisive contribution is called {\em phase ordering entropy}\/.
\end{enumerate}

\subsection{Mean phase}
\label{s2.2}
Admitting the values $\phi=0$ and $\phi=\pi/2$ in Eq.\ (\ref{e2}) means
that we use the functions $\sin(qx)$ and $\cos(qx)$. This is equivalent
to the use of $\exp({\rm i} k_1 x)$ and $\exp(-{\rm i} k_1 x)$, i.e.,
to the case (\ref{e1}) of no phase ordering. Using the single particle
functions (\ref{e2}) with one definite phase (lets say $\phi =0$)
effectively implies that only every second single particle state is
occupied. This can be accounted for by the substitutions $\sum_{\bf k}
\to (1/2)\sum_{\bf k}$ and $n_{\bf k}\to 2 n_{\bf k}$ in the
expression (\ref{e5}), i.e.,
\begin{equation}
\label{e6}
S_{\rm IBG}' = \frac{k_{\rm B}}{2} \sum_{{\bf k}}
\left[  (1 +  2 n_{\bf k}) \ln (1 +  2 n_{\bf k} )
 -  2 n_{\bf k} \ln (2n_{\bf k} )\right]
.
\end{equation}
The step from the entropy (\ref{e5}) to the entropy (\ref{e6}) takes
into account the choice $\phi=0$ for all atoms. Deviating from this, we
could admit different $\phi_q$ values in Eq.\ (\ref{e2}) for different
$q$'s without destroying the considered correlations. This is, however,
a negligible effect because the increase in entropy would be $\Delta
S/N=\sum_q\ldots /N \sim {\cal O}(N^{-2/3}) \approx 0$.

The expectation value of $S_{\rm IBG}'$ has similar temperature
dependence as that of $S_{\rm IBG}$ (Sec.\ \ref{s3.2}).

\subsection{Phase variances}
\label{s2.3}
Anderson \cite{an66} has established a connection between the condensate
wave function and {\em locally coherent states}\/. The coherent states
are constructed in suitably chosen subvolumes $\Delta V$. The volumes
must be large enough in order to define a mean phase $\bar\phi$ with a
small variance $\Delta \phi\ll 1$; this requires that the number of
atoms in $\Delta V$ is large compared to 1. On the other hand, the
volumes $\Delta V$ must be small enough in order to ensure $\bar\phi
\approx \mbox{const.}$ in spite of macroscopic perturbations or flows.

The connection between the condensate wave function and locally
coherent states may be carried over to noncondensed phase ordered
states (\ref{e2}) provided that $n_{\bf k}\gg 1$. This means that we
may construct coherent states that correspond to the single particle
functions (\ref{e2}). As in Anderson's work these coherent states are
restricted to finite volumes. For the following calculation it is not
necessary to quantify the size of these subvolumes. 

Appendix \ref{A} presents an explicit construction of coherent states
that correspond to phase ordered single particle functions (\ref{e2}).
Here we restrict ourselves to the presented argument based on the
analogy to known work \cite{an66}.

If the atoms with momentum $\bf k$ form a coherent state then their
phase variance \cite{ba89} reads
\begin{equation}
\label{e7}
\Delta\phi_{\bf k} = \frac{1}{\displaystyle 2\,\sqrt{\bar n_{\bf k}}} 
\,.
\end{equation}
The bar denotes the quantum mechanical average. 

In Eq.\ (\ref{e1}) we may admit arbitrary phases $\phi_j$ for each
atom, $\varphi_{\bf k} ({\bf r}_j) \propto  \exp (\,{\rm i}\,{\bf
k}\cdot{\bf r}_j + \phi_j)$. These phases $\phi_j$ are of no influence
on physical quantities (like the single particle kinetic energy or an
interaction matrix element). In particular, they are of no influence on
the statistical counting. Such phases are usually ignored.

Contrary to this, arbitrary phases $\phi_j$ in Eq.\ (\ref{e2}),
$\varphi^{\rm p.o.}_{\bf k} \propto \sin(q x_j + \phi_j)$, would
destroy the considered correlations. The phase ordering requires,
therefore, that these phases $\phi_j$ are adequately restricted, i.e.,
to an interval of the order (\ref{e7}) around a common mean value. A
random phase corresponds to a phase variance $\Delta \phi_{\rm ra} =
\pi /\sqrt{3}$.  (We use the phase definition by Barnett et
al. \cite{ba89}; the numerical factors are, however, without influence
on the critical part of the entropy.) Restricting the phase $\phi_j$ of
the $j$th atom to the interval (\ref{e7}) corresponds to a reduction
factor $\Delta\phi_{\bf k}/ \Delta\phi_{\rm ra}$. For all atoms this
leads to the entropy contribution
\begin{equation}
\label{e8} 
S_{\rm p.o.} = 
k_{\rm B} \ln \prod_{\bf k} 
\left(\frac{\Delta\phi_{\bf k}}{\Delta\phi_{\rm ra}}
\right)^{\!\mbox{\footnotesize$\bar n_{\bf k}$}} 
 \approx 
-\frac{k_{\rm B}}{2}\,\sum_{{\bf k}} \bar n_{\bf k}\,\ln \big( \bar
n_{\bf k}\big)  \qquad (\bar n_{\bf k} \gg 1) .
\end{equation}
This expression appears to be the most simple and plausible way of
accounting for the phase restrictions.

Phase ordering requires that the considered phases are reasonably well
defined, i.e., $\Delta\phi_{\bf k}\ll 1$ or $\bar n_{\bf k} \gg 1$. The
condition $\bar n_{\bf k} \gg 1$ has been used in the last step in
Eq.\ (\ref{e8}). We note that the critical terms in the expectation
value of the entropy are solely due to the contributions from $\bar
n_{\bf k} \gg 1$. The expectation value of $S_{\rm p.o.}$  will exhibit
a logarithmic singularity (Sec.\ \ref{s3.3}), i.e., the phase ordering
entropy is the decisive contribution in our model.

For $n_{\bf k} \gg 1$, expression (\ref{e5}) yields $S_{\rm IBG} \sim
\sum k_{\rm B} \ln n_{\bf k}$. From this and Eq. (\ref{e8}) it follows
that the low ${\bf k}$ contributions to the sum $S_{\rm IBG} +
S_{\rm p.o.}$ are negative which might seem disturbing. The entropy
must, however, not be attributed separately to each ${\bf k}$ level; it
rather results from the distribution of all particles onto the
available levels.  Consider, for example, an isolated level with the
single particle energy $\varepsilon_1$ that is occupied by $n_1$ atoms.
Then we may attribute the energy $\varepsilon_1 n_1$ to this level but
we cannot say that $k_{\rm B} \ln n_1$ is the entropy of this
(isolated) level.  Alternatively one may argue as follows: The low
${\bf k}$ contributions to $S_{\rm IBG}$ (or $S'_{\rm IBG}$) are small
and non\-critical.  Consequently, any entropy expression that leads to
the experimentally observed critical behavior (i.e., to $S_{\rm
exp}\propto -t \ln |t|$, where $t$ is the relative temperature) must
override the low ${\bf k}$ contribution of $S_{\rm IBG}$ (similarly as
it is done by $S_{\rm p.o.}$).

\section{Thermodynamic entropy}
\label{s3}
\subsection{Statistical assumptions}
\label{s3.1}
In order to determine the thermodynamic entropy we need the expectation
values $\langle n_{\bf k}\rangle$ of the occupation numbers. For the
wave function (\ref{e4}) with the single particle functions (\ref{e1})
and with hard sphere Jastrow factors, Chester  \cite{ch55} obtained the
IBG values, i.e., $\langle n_{\bf k}\rangle = \langle n_{\bf
k}\rangle_{\rm IBG}$. This means that the IBG occupation numbers
remain valid in spite of hard core interactions. The considered phase
ordering correlations are small in the sense that $|E_{\rm corr}|/N\ll
k_{\rm B}T_\lambda$.  Therefore, it appears to be a sensible
approximation to use the IBG occupation numbers for the phase ordered
states, too. The phenomenological assumption
\begin{equation}
\label{e9}
\langle n_{\bf k}\rangle = \langle n_{\bf k}\rangle_{\rm IBG}
\end{equation}
is the basis of our description of the phase transition. It implies
that we consider an {\em almost}\/ ideal Bose gas model. This
classification refers to the statistical assumption (\ref{e9})
but not to an assumption about weak interactions.

The thermodynamic entropy is obtained by
\begin{equation}
\label{e10}
S(T,V,N) = \langle S( n_{\bf k}) \rangle = S(\langle n_{\bf k}
\rangle) ,
\end{equation}
where
\begin{equation}
\label{e11}
S(n_{\bf k}) = S'_{\rm IBG} + S_{\rm p.o.} .
\end{equation}
The contributions $S'_{\rm IBG}$ and $S_{\rm p.o.}$ are defined in
Eqs.\ (\ref{e6}) and (\ref{e8}), respectively. For the relevant states
($n_{\bf k} \gg 1$) the quantum mechanical variances $\Delta n_{\bf
k}/\bar n_{\bf k} \sim 1/\sqrt{\bar n_{\bf k}}\ll 1$ are small compared
to the statistical variances.  Therefore, we omit a distinction between
$\langle \bar n_{\bf k}\rangle$ and $\langle n_{\bf k}\rangle$.

The average occupation numbers of the IBG are
\begin{equation}
\label{e12}
\langle n_{\bf k}\rangle =
\frac{1}{\exp \,[(\,\varepsilon_{\bf k} - \mu )/k_{\rm B}T] -1} =
\frac{1}{\,\exp \,(x^2 + \tau^2)-1} \,.
\end{equation}
Here $\varepsilon_{\bf k} = \hbar^2k^2/2m$ is the single particle
energy and $\mu$ is the chemical potential. We use the dimensionless
quantities $x$ and $\tau$,
\begin{equation}
\label{e13}
x^2 = \frac{\varepsilon_{\bf k} }{k_{\rm B} T} = \frac{\lambda^2
k^2}{4\pi}
\quad\mbox{and}\quad
\tau^2 = -\frac{\mu }{k_{\rm B} T}\,,
\end{equation}
where $\lambda = 2\pi \hbar /\sqrt{2\pi m k_{\rm B}T}$ denotes the
thermal wave length. The transition temperature $T_\lambda$ is given by
the condition
\begin{equation}
\label{e14}
\lambda (T_\lambda ) = \left[ \,v^{\,} \,\zeta (3/2)\,\right]^{1/3} ,
\end{equation}
where $\zeta (3/2) \approx 2.6124$ denotes Riemann's zeta function and
$v=V/N$ is the volume per particle. In the following we use the
relative temperature
\begin{equation}
\label{e15}
t = t(T,v) = \frac{T-T_\lambda }{T_\lambda } \,.
\end{equation}
The chemical potential $\mu$ vanishes at the transition point. For
$|t|\ll 1$ the particle number condition $N=\sum \langle n_{\bf
k}\rangle $ yields
\begin{equation}
\label{e16}
\tau (t) = \sqrt{\frac{-\mu}{k_{\rm B}T}} = 
\left\{ \begin{array}{ccl}
a\,t + b\,t^2 + \ldots       &\quad & (t>0) \\[1mm]
a'\, |t| +b\/' \, t^2+\ldots &      & (t<0)\,. \end{array} \right.
\end{equation}
The coefficients are known for the IBG, in particular $a = (3/4)\zeta
(3/2)/\sqrt{\pi}$ and $a'=b'=\ldots =0$. As a generalization of the IBG
we admit $\tau = a'\,|t|+ \ldots $ with $a'\ge 0$ for $t<0$.  This
generalization does not change the character of the transition (neither
the point of transition nor the critical exponents). Using the second
line in Eq.\ (\ref{e16}) and the particle number condition we obtain
the condensate fraction
\begin{equation}
\label{e17}
\frac{n_0(t)}{N} = \left( \frac{3}{2} + 
\frac{2\sqrt{\pi }\,a'}{\zeta (3/2)}\right) \, |t| + g\, t^2 +\ldots\
.
\end{equation}
Here $n_0(t)$ is the expectation value of the occupation number of the
lowest energy level. 

There are a number of reasons for generalizing the IBG by admitting
$\tau = a'\,|t| + \ldots $ with a coefficient $a'\ge 0$ in the
expansion (\ref{e16}): Formally, this generalization leads to a more
symmetric form of this expansion. Physically, a positive value of $a'$
implies a temperature dependent energy gap $\Delta \approx k_{\rm B} T
a'^2 t^2$ between the condensate and the noncondensed particles and
implies that the lowest level is more rapidly occupied than in the IBG.
Such a behavior leads to a better agreement between the calculated and
the experimental temperature dependence of various quantities, in
particular of the superfluid density \cite{fl99}. As a last argument we
mention that $a'> 0$ removes \cite{bl92} the unphysical divergence
($1/k^2$ for $k\to 0$) of the static structure function of the IBG for
$t<0$. Our main result will not depend on the parameter $a'$.

The well-known expectation value of the IBG entropy (\ref{e5}) is given
by
\begin{equation}
\label{e18}
\frac{S_{\rm IBG}(T,V,N)}{Nk_{\rm B}} = 
\frac{5}{2} \frac{v}{\lambda^3} \, g_{5/2}(\tau) + \tau^2 ,
\end{equation}
where
\begin{equation}
\label{e19}
g_{\nu}(\tau) 
= \sum_{n=1}^{\infty} \frac{\exp(-n\tau^2)}{n^\nu}
= \sum_{n=1}^{\infty} \frac{z^n}{n^\nu}
\end{equation}
defines Riemann's generalized zeta function. Usually $z =
\exp(\beta\mu) = \exp(-\tau^2) $ is taken as the argument of this
function. We prefer the argument $\tau$ because of its close relation
($\tau \sim |t|$) to the relative temperature.

For the IBG entropy one has to use $\tau \equiv 0$ in Eq.\ (\ref{e18})
for $t<0$. The modification $\tau = a' |t|$ with $a'> 0$ leads to a
specific heat that is continuous at $T_\lambda$ (as in the pure IBG)
but that falls off more rapidly for $t<0$.

\subsection{Mean phase contribution}
\label{s3.2}
Because of the common mean phases the IBG expression (\ref{e5})
is replaced by $S'_{\rm IBG}$, Eq.\ (\ref{e6}). The expectation value of
the entropy $S'_{\rm IBG}$ is
\begin{equation}
\label{e20} 
\frac{S_{\rm IBG}'(T,V,N)}{Nk_{\rm B}} = \frac{1}{2} \sum_{{\bf k}}
\Big[  (1 +  2 \langle n_{\bf k}\rangle )\,
\ln (1 +  2 \langle n_{\bf k}\rangle ) 
 - 2\langle n_{\bf k}\rangle\,\ln (2\langle n_{\bf k}\rangle )
\Big] .
\end{equation}
For a first, crude estimate we take into account only those
contributions that come from $\langle n_{\bf k}\rangle  \gg 1$.
This yields $S_{\rm IBG} \sim k_{\rm B} \sum_{\bf k} \ln\langle n_{\bf
k}\rangle $ and $S_{\rm IBG}'
\sim (k_{\rm B}/2) \sum_{\bf k} \ln \langle n_{\bf k}\rangle $, i.e.,
$S'_{\rm IBG}(T,V,N)\sim S_{\rm IBG}(T,V,N)/2$.

The actual results for $S_{\rm IBG}$ and $S'_{\rm IBG}$ near
the transition point are
\begin{eqnarray}
S_{\rm IBG}(T,V,N) &\approx & 1.28\, Nk_{\rm B} 
\left(1 + \frac{3}{2}\, t + {\cal O}(t^2)\right),
\label{e21} 
\\
S'_{\rm IBG}(T,V,N) &\approx & 0.96\, N k_{\rm B} 
\left(1 + \frac{3}{2}\, t + {\cal O}(t^2)\right).
\label{e22} 
\end{eqnarray}
Eq.\ (\ref{e21}) follows from the expression (\ref{e18}).  The value of
$S'_{\rm IBG}$ at the lambda point has been determined by a numerical
integration of the r.h.s.\ of Eq.\ (\ref{e20}). The next term in the
expansion is due to the prefactor $v/\lambda^3\propto T^{3/2}$ in Eq.\
(\ref{e18}). This prefactor originates from the replacement of the
momentum sum by an integral over dimensionless variables; it occurs in
both entropy expressions in the same way. The higher terms ${\cal
O}(t^2)$ in Eqs.\ (\ref{e21}) and (\ref{e22}) do not agree; moreover,
the coefficients of these $t^2$ terms are different for $t>0$ and
$t<0$.

\subsection{Phase variance contribution}
\label{s3.3}
The expectation value of the phase ordering entropy $S_{\rm p.o.}$,
Eq.\ (\ref{e8}), reads
\begin{equation}
\label{e23}
 \frac{S_{\rm p.o.}(T,V,N)}{Nk_{\rm B}}
 =   -\frac{1}{2N}\, \sum_{\bf k}  \langle n_{\bf k} \rangle
\,\ln \langle n_{\bf k}\rangle \qquad (|t| < 0.1).
\end{equation}
For the phase ordering we required $\Delta \phi_{\bf k}\ll 1$ in Eq.\
(\ref{e8}), lets say $\Delta \phi_{\bf k} < 0.1$ or $\langle n_{\bf k}
\rangle > 10^{2}$. Because of $\langle n_{\bf k} \rangle \approx
1/(x^2+\tau^2)$ this implies $x\lesssim 0.1$ and $\tau\lesssim 0.1$ or
$|t| < 0.1$. The condition $|t| < 0.1$ indicates the temperature range
in which phase ordering and the corresponding entropy contribution are
expected to be relevant.

For evaluating the entropy (\ref{e23}) the momentum sum is replaced by
an integral,
\begin{equation}
\label{e24}
\sum_{\bf k} \ldots  \longrightarrow 
 \frac{V}{(2\pi)^3}\int_0^\infty \! dk\, 4\pi k^2 \ldots\ .
\end{equation}
This step has to be accompanied by a separate consideration of a
potential condensate fraction; this will be done below in Eq.\
(\ref{e28}).

For a subvolume of size $V_0$ the r.h.s. of Eq.\ (\ref{e24}) would be
$V_0/(2\pi)^3\int_0^\infty \! dk\, 4\pi k^2$. The subsequent summation
over all subvolumes yields the factor $V/V_0$; this summation has been
taken into account in the prefactor on the r.h.s.\ of Eq.\ (\ref{e24}).
In Appendix \ref{B} we discuss the validity of the step (\ref{e24})
with respect to finite spacing $\Delta k= \pi/V_0^{1/3}$ of the
momentum values.

We insert the replacement (\ref{e24}), the dimensionless quantities
(\ref{e13}) and the occupation numbers (\ref{e12}) into Eq.\
(\ref{e23}):
\begin{equation}
\label{e25}
 \frac{S_{\rm p.o.}}{Nk_{\rm B}} =
\frac{2}{\sqrt{\pi}} \frac{v}{\lambda^3}
\int_0^\infty \! dx \,x^2\,
\frac{\ln [\exp(x^2 + \tau^2)-1]}{\exp(x^2 + \tau^2)-1} 
= \frac{2}{\sqrt{\pi}} \frac{v}{\lambda^3}\, J(\tau) .
\end{equation}
The critical $\tau$ dependence of the integral $J(\tau)$ can be
determined analytically (App.\ \ref{C}),
\begin{equation}
\label{e26}
J(\tau) 
= J(0) - \pi \tau \ln\tau + \pi \tau [1-\ln(2)] + {\cal O}(\tau^2).
\end{equation}
Using this, $\tau =  \big(3\, \zeta (3/2)/4 \sqrt{\pi}\, \big)\,
|t|+  {\cal O}(t^2) $, and $v/\lambda^3 = 1/\zeta(3/2) + {\cal O}(t)$
we obtain
\begin{equation}
\label{e27}
S_{\rm p.o.}(T,V,N) = - \frac{3 N  k_{\rm B}}{2}\, t  \ln |t|\,
\pm \ldots \qquad (|t| < 0.1)
\end{equation}
for the leading singularity of the phase ordering entropy for $t>0$.
As we will see, this result holds for $t<0$, too.

For $t<0$ the prescription (\ref{e24}) has to be accompanied by the
replacement 
\begin{equation}
\label{e28}
\langle n_{\bf k}\rangle  \to \langle n_{\bf k}\rangle  + n_0(t)
\, \delta({\bf k})  .
\end{equation}
A phase variance $\Delta\phi_0\sim 1/\sqrt{n_0}\approx 0$ stands for a
macroscopic phase coherence. This is appropriate for a superfluid flow
but not for the considered local phase ordering. For the phase
variances (\ref{e7}) we use, therefore, the continuous part of the
occupation numbers $\langle n_{\bf k}\rangle$ only. The contribution
$S_{\rm p.o.}^{\rm cond}$ in Eq.\ (\ref{e23}) due to the condensed
particles reads then
\begin{equation}
\label{e29}
\frac{S_{\rm p.o.}^{\rm cond}}{Nk_{\rm B}} =
- \frac{n_0(t)}{2N}\, 
\ln \langle n_{{\bf k}\to 0} \rangle  = 
 \frac{n_0(t)}{2N}\, \ln[\exp(\tau^2)-1] \,.
\end{equation}
This can be evaluated by using Eq.\ (\ref{e17}) for $n_0(t)/N$ and
Eq.\ (\ref{e16}) for $\tau (t)$.

For $t<0$ the contributions (\ref{e25}) and (\ref{e29}) have to be
added. The $a'$ term of $n_0(t)/N = [3/2 + 2\sqrt{\pi} a'/\zeta(3/3)]
|t| + \ldots\ $ in Eq.\ (\ref{e29}) cancels exactly the logarithmic
term in Eq.\ (\ref{e25}).
The surviving leading term comes from $n_0(t)/N
= (3/2) |t| +\dots\ $ and $\ln[\exp(\tau^2)-1] = 2\ln|t| +\ldots\ $
yielding again Eq.\ (\ref{e27}).

\subsection{Specific heat}
The specific heat per particle may be written as 
\begin{equation}
\label{e30}
c_V = \frac{1}{N}\left(\frac{\partial S}{\partial t}\right)_{\!V,N}
= \left\{\begin{array}{lll} -A\ln |t| + B + \ldots && (t>0)
\\[1mm]
 -A'\ln |t| + B' + \ldots && (t <0)\,.
\end{array}\right.
\end{equation}
From Eq.\ (\ref{e27}) we obtain the theoretical amplitudes
\begin{equation}
\label{e31}
A_{\rm theor} = A_{\rm theor}' = \frac{3}{2}\, k_{\rm B}.
\end{equation}
If the experimental data for $c_V$ are fitted by the form (\ref{e30})
one finds $A_{\rm exp} \approx 0.63 \, k_{\rm B}$ and $A_{\rm exp}'
\approx 0.59 \, k_{\rm B}$; these values are taken from Eq.\ (4.2) (for
low pressure) of Ref.\  \cite{ah73}. It is remarkable that we 
obtain a parameter free result and a sensible size for these
amplitudes. The experimental ratio $A_{\rm exp} /A_{\rm exp}'$ deviates
from 1 by a few percent.  In our model, the contribution (\ref{e25}) of
the noncondensed particles to $S_{\rm p.o.}$ has different signs above
and below the lambda point. A correction in this contribution could,
therefore, lead to a deviation from $A_{\rm theor} /A_{\rm theor}' =
1$.

Experimentally the specific heat $c_P$ at constant pressure is more
readily accessible. It is this quantity that has been measured
\cite{gr73,pa82,li83,li96} over several decades of the relative
temperature, eventually down to $|t| = 2\times 10^{-9}$ in a
microgravity experiment \cite{li96}. In the context of these
measurements and their analyses we note the following points:
\begin{enumerate}
\item
Theoretically, a logarithmic behavior of one of the two quantities,
$c_P$ or $c_V$, implies \cite{le67,kr89} an analogous behavior of the
other quantity in the experimentally accessible range. The
statement \cite{bu61} that only one of these quantities may
diverge for $|t|\to 0$ applies to very low (experimentally
inaccessible) $|t|$ values only \cite{kr89}.

The experimental values for $c_P$ may be fitted by the form
(\ref{e30}), too. The corresponding coefficients (as given by Eq.\ (47)
of Ref.\  \cite{gr73}) differ from that for $c_V$ by about 5 to
10\%{}.  For comparing our result (\ref{e30}) with the experiment we
may, therefore, also consider the experimental $c_P$.
\item
If the critical exponents $\alpha$ and $\alpha'$ are used as fit
parameters one finds \cite{gr73,li83,li96} small negative values.
The following points indicate that this does not rule out a logarithmic
singularity (corresponding to $\alpha \to 0$ and $\alpha' \to 0$) for
the leading term:
\begin{enumerate}
\item
Arp \cite{ar90} fitted the data of many experiments in order to obtain
an equation of state for helium. He considered specifically the
question of the critical exponent $\alpha$ (Sec.\ 13.1.1 of Ref.\
 \cite{ar90}) and found no statistical significance for a
deviation from a leading logarithmic form. Consequently, he adopted the
value $\alpha = \alpha' = 0$ (standing for a logarithmic function) in
his expression for the specific heat $c_V$.
\item
By the scaling law $\alpha' = 2- 3\nu$ the critical exponent for the
specific heat for $t<0$ is related to that of the superfluid density
$\varrho_{\rm s} \sim |t|^\nu$. The deviations from $\alpha'=0$ and
$\nu=2/3$ found in standard fits are consistent \cite{li96} with this
scaling law.

In Ref.\  \cite{fl99} we presented an alternative fit formula for
the superfluid density in which the leading exponent $\nu$ equals
exactly 2/3. This fit formula reproduced the data better than standard
fit formulas. This means that the value $\nu =2/3$ (corresponding to
$\alpha'=0$) is not ruled out by the experiment, and that the value of
the leading exponent (found in a fit) might depend on the higher order
terms (used in the fit).
\end{enumerate}
\end{enumerate}
In view of these points, we consider it to be an open question whether
the true behavior deviates from a logarithmic one in a way that cannot
be accounted for by higher order terms.

We turn now to the coefficients $B$ and $B'$ of the specific heat
(\ref{e30}). From $S_{\rm IBG}'$, Eq.\ (\ref{e22}), we get a
contribution of $1.44\, k_{\rm B}$ for both, $B_{\rm theor}$ and
$B'_{\rm theor}$. In Eq.\ (\ref{e27}) we add the terms that are linear
in $t$; these terms follow from Eqs.\ (\ref{e25}) and (\ref{e29}).
Including all contributions we obtain
\begin{equation}
\label{e32}
B_{\rm theor} \approx -0.52 \, k_{\rm B}, \qquad 
B_{\rm theor}' = -3.7\, k_{\rm B}.
\end{equation}
For $B_{\rm theor}'$ we used the parameter value $a'\approx 3$ that is
(rather uniquely) determined \cite{fl99} by adjusting the model
expression for the superfluid density to the experimental temperature
dependence.

The value for $B_{\rm theor}$ compares reasonably well with the
experimental value $B_{\rm exp} \approx -0.84 \, k_{\rm B}$ (from Eq.\
(4.5) of Ref.\  \cite{ah73} for low pressure).

There is a large discrepancy between $B_{\rm theor}'$ and the
experimental value \cite{ah73} $B'_{\rm exp}\approx 2.0\,k_{\rm B}$.
In the IBG, the coefficient of the $t^2$ term in the free energy
$F_{\rm IBG}$ is the same above and below the transition. This means
that the IBG does not reproduce the jump of the specific heat that
normally accompanies the occurrence of a macroscopic order parameter
field. For a crude estimate of this missing jump we consider the Landau
free energy $F_{\rm Landau}/(Nk_{\rm B}T_\lambda) = r t|\psi|^2 + u
|\psi|^4 $. This Landau model yields a jump of the specific heat,
$\Delta c_V /k_{\rm B} = r^2/2u$.  The identification $\langle
|\psi|^2\rangle = n_0(t)/N$ relates the coefficient in $\langle
|\psi|^2\rangle = (r/2u) |t|$ to that in Eq.\ (\ref{e17}).  This
connection leads to $\Delta c_V = r\, [3/2 + 2\sqrt{\pi}a'/
\zeta(3/2)]$. Using again $a'\approx 3$ yields $\Delta c_V \approx
5.6\, r$. For the parameter $r$ in the expression for $F_{\rm
Landau}/(Nk_{\rm B} T_\lambda)$ we may expect $r\sim {\cal O}(1)$. The
resulting $\Delta c_V\sim 5.6\, k_{\rm B}$ could close the gap
between $B'_{\rm theor}$ of Eq.\ (\ref{e32}) and $B'_{\rm exp}$.

\subsection{Entropy at the lambda point}
As a last point we consider the value of the entropy at the
lambda point. Evaluating the phase ordering entropy (\ref{e25}) as
it stands yields
\begin{equation}
\label{e33}
S_{\rm p.o.}(T_\lambda) = \frac{2 J(0)}{\sqrt{\pi}\, \zeta(3/2)} 
\,N k_{\rm B} \approx -0.51 \,N k_{\rm B} .
\end{equation}
This result contains, however, contributions from momenta for which the
condition $\langle n_{\bf k}\rangle \gg 1$ is not satisfied.

We estimate the value $S^{\rm est}_{\rm p.o.}(T_\lambda)$ that comes
from the contributions with $\langle n_{\bf k}\rangle \gg 1$ alone.
For small and positive $t$ values we may assume the form $S_{\rm
p.o.}/N \approx S^{\rm est}_{\rm p.o.}(T_\lambda)/N - A_{\rm theor} \,
t \ln(t) + (B_{\rm p.o.} + A_{\rm theor}) \, t$.  Following Eq.\
(\ref{e23}) we argued that the phase ordering entropy should be
relevant in the range $|t|< 0.1$ only. This implies $S_{\rm
p.o.}(t=0.1)\approx 0$ leading to $S_{\rm p.o.}^{\rm est}(T_\lambda)
\approx -0.3\,N k_{\rm B}$.

We conclude that the result (\ref{e33}) might be too large by roughly a
factor of 2. One may, therefore, expect that the contribution $S_{\rm
p.o.}(T_\lambda)$ accounts for the difference between $S'_{\rm IBG}
\approx 0.96\,Nk_{\rm B}$ and the experimental value \cite{si84} $S_{\rm
exp}(T_\lambda) \approx 0.76\,Nk_{\rm B}$.

\section{Concluding remarks}
We summarize the main features of the presented phenomenological many-body
model:
\begin{enumerate}
\item
Following Chester \cite{ch55} we use the IBG many-body states multiplied
by Jastrow factors.
\item
We assume that local phase correlations are the relevant correlations
near the lambda point. These correlations are specified by the single
particle functions (\ref{e2}).
\item
The local phase ordering can be described by coherent states. The phase
variances of these coherent states are the basis for the statistical
counting of the phase restrictions. This leads to the expression
(\ref{e8}) for the phase ordering entropy.
\item
For evaluating the phase ordering entropy we use the IBG expression for
the average occupation numbers.
\end{enumerate}
It is clear that global phase ordering plays a decisive role in liquid
helium below the transition point. This makes our assumption that local
phase ordering are relevant correlations near the lambda point to some
extent plausible. We have proposed a specific kind of the local phase
ordering in the framework of an almost ideal Bose gas model. The use of
the IBG occupation numbers is a phenomenological assumption.

Our model yields a logarithmic singularity of the specific heat in a
straightforward and rather simple way: The expression (\ref{e8}) for
the phase ordering entropy follows from the phase variances (\ref{e7}).
Evaluating this expression with the IBG occupation number yields the
logarithmic singularity (\ref{e27}) (immediately for $t>0$, and after a
slight generalization of the occupation numbers for $t<0$).

We summarize the main results of our almost ideal Bose gas model:
\begin{enumerate}
\item
The model yields a realistic expression for the specific heat of liquid
helium at the $\lambda$ transition. For the strength of the logarithmic
singularity the remarkable result $A=A'= 3\,k_{\rm B}/2$ is obtained.
\item
The model retains the essential features of the IBG. Therefore, it
strengthens the suggested relation \cite{lo54,fe53} between the
Bose-Einstein condensation and the $\lambda$ transition.
\item
The model offers an intriguing picture for the relevant correlations
near the $\lambda$ point. This picture provides also perspectives
beyond the specific heat (some of which are indicated in App.\
\ref{B}).
\end{enumerate}
Various other quantities may be calculated in the framework of our
almost ideal Bose gas model: In an earlier paper \cite{fl91} we
calculated the energy as a function of the temperature; the present
derivation of the specific heat appears to be more direct. The critical
exponent $\beta=1/2$ of the condensate fraction may be reconciled with
the actual behavior of the superfluid density by assuming that
noncondensed particles move coherently with the condensate \cite{fl99}.
This idea leads to observable consequences that have been discussed in
Ref.\ \cite{sc94}.

\clearpage

\begin{appendix}
\section{Coherent states}
\label{A}
In order to justify the use of relation (\ref{e7}) we establish a
connection between the phase ordered single particle functions
(\ref{e2}) and coherent states. Our procedure is analogous to
Anderson's construction \cite{an66} of coherent states for the
condensate.

The many-body wave function (\ref{e4}) can be written in the form
\begin{equation}
\label{a1}
\Psi = {\cal S} F 
\prod_{j=1}^{n_0} \varphi_0({\bf r}_j)
\prod_{j=n_0+1}^{n_0+n_1} \varphi_1 ({\bf r}_j)
\cdot \ldots
\, .
\end{equation}
The  single particle functions (phase ordered or not) are denoted by
$\varphi_i$, where $i=0,1,2,\ldots$ follows the energy sequence.
Without the Jastrow factor the off-diagonal single particle density
reads
\begin{equation}
\label{a2}
\varrho({\bf r},{\bf r}') =
\left\langle \Psi 
\left|\hat\psi^+({\bf r}')\, \hat\psi({\bf r}) \right|\Psi 
\right\rangle  =
  n_0\, \varphi_0 ({\bf r})^*  \varphi_0 ({\bf r}')
+ n_1\, \varphi_1 ({\bf r})^*  \varphi_1 ({\bf r})  + \ldots\ ,
\end{equation}
where $\hat\psi^+$ and $\hat\psi$ are the particle creation and
annihilation operator, respectively. Including the Jastrow factor the
condensate contribution in Eq.\ (\ref{a2}) is depleted \cite{pe56,wi87}
from $n_0$ to a lower value $n_{\rm c}$. Similar effects are to be
expected for next low-momentum contributions.

All terms in Eq.\ (\ref{a2}) besides the first term vanish for $|{\bf
r}-{\bf r}'|\to \infty$. This {\em off-diagonal long range
order}\/ \cite{pe56} is considered \cite{no90} as the decisive criterium
for superfluidity.

Anderson \cite{an66} related the off-diagonal long range order to the
mean field aspect by constructing localized coherent states. The
coherent states $|{\rm coh}\rangle$ are defined such that $\langle {\rm
coh}| \hat\psi |{\rm coh}\rangle \propto \exp({\rm i} \phi )$, where
$\phi$ is the phase of the condensate wave function $\varphi_0\propto
\exp({\rm i} \phi )$. Accordingly, we will construct coherent states $
|{\rm coh}\rangle$ for which $\langle {\rm coh}| \hat\psi |{\rm
coh}\rangle $ is proportional to the phase ordered single particle
function (\ref{e2}).  Anderson's aim was to show that the phase of the
condensate and the corresponding particle number may be treated as
macroscopic variables. Our aim is to justify Eq.\ (\ref{e7}) by
relating the wave functions (\ref{e2}) to coherent states.

Anderson \cite{an66} introduced finite volumes $\Delta V$ in which the
phase $\phi$ of the condensate wave function $\varphi_0$ is
approximately constant. We consider phase ordered single particle
functions that are restricted to finite subvolumes of
size $V_0$. In the following we identify $\Delta V$ with $V_0$ and
treat both cases simultaneously.

The single particle functions $\varphi_{k}$ shall form an orthonormal
set in one subvolume of size $V_0$. Then we may write down the
following relations between the field operator $\hat \psi({\bf r})$ and
the annihilation operator $\hat c_{\bf k}$:
\begin{eqnarray}
\label{a3}
\hat \psi ({\bf r}) & =&  \sum_{\bf k} \varphi_{\bf k}({\bf r})\,
\hat c_{\bf k} \,,
\\
\label{a4}
\hat c_{\bf k} & =&  \int_{V_0}\! d^3r \, \varphi^*_{\bf k}({\bf r})\,
\hat \psi ({\bf r}).
\end{eqnarray}
A state with $n_{\bf k}$ particles of momentum ${\bf k}$ is given by
\begin{equation}
\label{a5}
|n \rangle = |n_{\bf k}\rangle = \frac{1}{\sqrt{n_{\bf k}!}}\, 
\hat c^+_{\bf k}\, |{\rm vac}\rangle ,
\end{equation}
where $|{\rm vac}\rangle$ is the vacuum state. In the following we
restrict ourselves to one specific momentum and omit the index ${\bf
k}$. The coherent state \cite{zh90} is constructed as
\begin{equation}
\label{a6}
|{\rm coh} \rangle = \exp(-|z|^2/2) \sum_{n_=0}^\infty
 \frac{z^{n}}{\sqrt{n!}}\, |n \rangle .
\end{equation}
This state depends on a complex number $z$. Because of $\bar n =
\langle {\rm coh} | \hat c^+_{\bf k} \hat c_{\bf k} |{\rm coh} \rangle
= |z|^2$ we may set
\begin{equation}
\label{a7}
z = \sqrt{\bar n}\, \exp({\rm i} \phi ).
\end{equation}
For the state (\ref{a6}) with Eqs.\ (\ref{a5}) and (\ref{a7}) we find
\begin{equation}
\label{a8}
\langle {\rm coh} | \hat\psi ({\bf r}) |{\rm coh} \rangle =
 \sqrt{\bar n }\,  \exp({\rm i} \phi )
 \, \varphi_{\bf k}({\bf r}) 
\end{equation}
for the expectation value of the field operator.

Using
\begin{equation}
\label{a9}
\varphi_{0}({\bf r}) =\left\{\begin{array}{lll}
1/\sqrt{V_0} &\,\,& ({\bf r} \in V_0)\\
 0 && ({\bf r} \notin V_0)\,, \end{array}\right. 
\end{equation}
for the lowest state, Eq.\ (\ref{a8}) becomes $\langle {\rm coh}|
\hat\psi |{\rm coh}
\rangle = \sqrt{\bar n/V_0}\, \exp({\rm i} \phi )$. So far we have
reproduced Anderson's construction.

We are now going to connect the phase ordered single particle functions
(\ref{e2}) with coherent states. For this purpose we introduce the
single particle functions
\begin{eqnarray}
\label{a10}
\varphi_{{\bf k},+} ({\bf r}) & \propto &
 \exp\big[\,{\rm i}\,(k_2 y + k_3 z)\big]\, 
\exp(+{\rm i}q x )
\\[1mm]
\label{a11}
\varphi_{{\bf k},-} ({\bf r}) & \propto &
 \exp\big[\,{\rm i}\,(k_2 y + k_3 z)\big]\, 
\exp(-{\rm i}q x )
\end{eqnarray}
These functions shall be orthonormalized in the considered subvolume.
Analogously to Eqs.\ (\ref{a5}) with (\ref{a4}) we introduce the
$n$-particle states $|n\rangle_+$ and $|n\rangle_-$ that correspond to
$\varphi_{{\bf k},+}$ and $\varphi_{{\bf k},-}$, respectively. The
coherent state
\begin{equation}
\label{a12}
|{\rm coh} \rangle = \frac{\exp(-|z|^2/2)}{\sqrt{2}} \sum_{n =0}^\infty
 \frac{z^{n} |n \rangle_+ - z^{*n} |n \rangle_- }{\sqrt{n!} }
\end{equation}
with $z= \sqrt{\bar n }\exp({\rm i}\phi )$ yields then
\begin{equation}
\label{a13}
\langle {\rm coh} | \hat\psi ({\bf r}) |{\rm coh} \rangle
= \sqrt{\bar n} \,  \varphi^{\rm p.o.}_{\bf k}({\bf r}).
\end{equation}
The phase variance of the coherent state (\ref{a12}) is given by
\begin{equation}
\label{a14}
\Delta\phi_{\bf k} = \frac{1}{\displaystyle 2\,\sqrt{\bar n_{\bf k}}}.
\end{equation}
Here we have attached the index $\bf k$ again; the above discussion
referred to a single $\bf k$ value. Eq.\ (\ref{a14}) is a well-known
result \cite{ba89} for a coherent state of the standard form (\ref{a6}).
For the state (\ref{a12}) we may write $|{\rm coh} \rangle = (|{\rm
coh} \rangle_+ - |{\rm coh} \rangle_-)/\sqrt{2}$ in an obvious
notation. The phase operator acts separately within each of the two
sets of states, $\{ |n \rangle_+ \}$ or $\{ |n \rangle_- \}$. Moreover,
the states of one of these sets are orthogonal to that of the other set
(because of the functions (\ref{a10}) and (\ref{a11}) are mutually
orthogonal). Therefore, the phase variance of the state (\ref{a12}) is
one half of the sum of the phase variances of the states $|{\rm coh}
\rangle_+ $ and $|{\rm coh} \rangle_- $ that are of the standard form
(\ref{a6}). This means that we obtain the standard result (\ref{a14})
for our somewhat special coherent state (\ref{a12}), too.

The reason for establishing the relation between the phase ordered
single particle functions (\ref{e2}) and coherent states (\ref{a12}) is
the necessity to justify the relation (\ref{a14}) or, equivalently,
Eq.\ (\ref{e7}). The relation (\ref{e7}) is the basis for the phase
ordering entropy (\ref{e8}).

Instead of $|{\rm coh} \rangle = (|{\rm coh} \rangle_+ - |{\rm coh}
\rangle_-)/\sqrt{2}$ we may also consider the coherent
state $(|{\rm coh} \rangle_+ + |{\rm coh} \rangle_-)/\sqrt{2}$. This
orthogonal state corresponds to a cosinus in Eq.\ (\ref{e2}) instead of
a sinus. The considered correlation effect requires that only one of
these sets of states is occupied. This $2:1$ reduction has been
discussed in Section \ref{s2.2}.

With respect to the condition $n_{\bf k}\gg 1$ we note that the average
occupation numbers $\langle n_{\bf k}\rangle$, Eq.\ (\ref{e12}), do not
depend on the size $V_0$ of the subvolumes. The finite size implies,
however, a finite spacing of the momentum values (App.\ \ref{B}) that
takes care of the finite number $N_0=V_0/v$ of atoms in one subvolume.

\section{Phase coherence volume}
\label{B}
We start by explaining why finite phase correlations require that the
single particle functions (\ref{e2}) have to be localized within finite
volumes. Then we show that the lowest single particle functions
may extend over several of these subvolumes. This leads to the notion
of a phase coherence volume for which we obtain $V_{\rm coh} \sim
V_0/|t|^2$, where $V_0$ is the size of one subvolume. We discuss this
result in a number of points.

Two phase ordered states (\ref{e2}) with momentum ${\bf k}$ and ${\bf
k}'$ are correlated if $k_1=k_1'$. The restriction $k_1=k_1'$ cancels
one momentum sum in the expression for the energy. A momentum sum
$\sum_k \to \int\!dk/\Delta k$ is proportional to $1/\Delta k
=V^{1/3}/2\pi$ or to $N^{1/3}$. The correlation energy $E_{\rm corr}$
is, therefore, proportional to $N^{2/3}$ only. This is consistent with
the observation that a phase ordering of the kind (\ref{e2}) can be
obtained by physical boundary conditions at the wall of the considered
volume.  For an infinite volume such a surface effect vanishes ($E_{\rm
corr}/N\propto 1/N^{1/3} \to 0$). A finite correlation effect may,
however, be obtained by dividing the total volume $V$ into $V/V_0$
subvolumes of size $V_0$ and by requiring physical boundary conditions
at the walls of these subvolumes. For the whole system (including a
factor $V/V_0$ for summing over all subvolumes) we obtain then a result
of the form
\begin{equation}
\label{b1}
\frac{E_{\rm corr}}{N} = \frac{w_0}{N_0^{1/3}} \,,
\end{equation}
where $w_0<0$ is some average strength of the attractive part of the
interaction (for a more detailed discussion see Ref.\
 \cite{fl91}). This result shows that a finite correlation effect
requires a finite size $V_0 = N_0\, v$ of the subvolumes.

Finite volumes $V_0$ imply a finite spacing $\Delta k = \pi/V_0^{1/3}$
of the momentum values. Alternatively one may start from a finite
spacing and admit only the momentum values
\begin{equation}
\label{b2}
q_n = q_0 + n \cdot \Delta k, \qquad (n=0,1,2,\ldots),
\end{equation}
in the single particle functions (\ref{e2}). The spacing (\ref{b2})
leads to a finite correlation energy even if the single particle
functions are not localized. In a macroscopic system the possible $q$
values are, however, dense, and the entropy drives the particles into
the occupation of all available states. This is the reason why the finite
spacing (\ref{b2}) can be realized only for single particle functions
localized within subvolumes of the size $V_0=(\pi /\Delta k)^3$. In
this sense, a finite spacing implies finite volumes.

This discussion shows that we may either start from finite volumes or
from a finite $\Delta k$ value. There is, however, the following
difference: Starting from the finite spacing (\ref{b2}) we may admit a
$q_0$ value {\em below}\/ $\Delta k$ without damaging the correlation
energy. A single particle function with $q_0 < \Delta k$ can, however,
not be localized within $V_0$.

In view of this observation we consider the following modified picture.
Only the single particle functions (\ref{e2}) with $q_{n\ge 1}$ are
localized within $V_0$, the single particle function $\varphi_0$ with
$q_0$ may have a larger extension. We present an estimate for the
volume $V_{\rm coh}$ of the lowest single particle function
$\varphi_0$: Let $V_{\rm coh}$ be some multiple of $V_0$, that means
$V_{\rm coh} = W V_0$. In this case, there are $W$ single particle
states with $q < \Delta k$ in the volume $V_{\rm coh}$ out of which
only one (the one with $q_0$) is occupied. A redistribution of $n_0$
atoms over these $W$ states would increase the entropy by $\Delta s
\sim k_{\rm B}\ln (n_0)^W$. At the same time these atoms would loose
their correlation energy, $\Delta e_{\rm corr} \sim - n_0 w_0/
N_0^{1/3}$. The stability condition $T_\lambda\,\Delta s < \Delta
e_{\rm corr}$ yields the upper bound $W < [-w_0/(N_0^{1/3} k_{\rm B}
T_\lambda)]\, n_0 /\ln n_0 \sim n_0$.  Using $n_0 = \langle n_{{\bf
k}\to 0}\rangle = \tau^{-2} \sim t^{-2}$ leads to
\begin{equation}
\label{b3}
V_{\rm coh}(t)= W\,V_0 \sim \frac{V_0}{t^2} \,.
\end{equation}
We discuss this result in a number of points:
\begin{enumerate}
\item 
The lowest single particle functions $\varphi_0$ extend over volumes of
the size $V_{\rm coh}$. Within a volume $V_{\rm coh}$ the function
$\varphi_0$ defines the direction of the phase ordering ($x$ direction
in Eq.\ (\ref{e2})) and the phase $\phi$. Therefore, $V_{\rm coh}(t)$
of Eq.\ (\ref{b3}) constitutes a {\em phase coherence volume}\/.
\item
Strictly finite volumes for all particles would mean that the lower
bound of the integral $J(\tau)$ in Eq.\ (\ref{e25}) would be $\Delta x$
instead of zero. This would imply a cut of the logarithmic singularity.
The replacement (\ref{e24}) may, however, be justified by a lower bound
$q_0\sim 1/V_{\rm coh}$ and an average over somewhat different $q_0$
and $\Delta k$ values in different parts of the macroscopic system.
A closer examination of this point might lead to a modification of the
logarithmic singularity at very low $|t|$ values.

\item
In the estimate leading to Eq.\ (\ref{b3}) we used the continuous part
of the occupation number $\langle n_{{\bf k}\to 0}\rangle$ also for
$t<0$. Using instead the condensate occupation number $n_0(t)$ leads to
an infinite volume, $V'_{\rm coh}= \infty$. This is the adequate phase
coherence volume for the potentially macroscopic range of a superfluid
flow.

\item
Different phase coherence volumes of size $V_{\rm coh}$ within the
macroscopic system will, in general, correspond to different phase
directions ($x$ direction in Eq.\ (\ref{e2})). A specific phase
direction implies a small anisotropy of the static structure function.
In principle, the volumes $V_{\rm coh}$ and their increasing size for
$t\to 0$ should, therefore, be observable.
\item
When approaching the lambda point from above the phase coherence
volumes grow according to Eq.\ (\ref{b3}). At the lambda point the
coherence volume becomes infinite.  This picture has some similarity
with a ferromagnetic system.

Below the lambda point the finite volume (\ref{b3}) refers to the phase
ordering of real single particle functions of the form (\ref{e2}). At
the same time there is a potentially infinite range of phase coherence
of a superfluid current.  The corresponding phase coherence volume
$V_{\rm coh}'=\infty$ refers to the complex condensate wave $\varphi_0
\propto \exp[\,{\rm i}\,\phi({\bf r})]$.

\item 
Adjusting the correlation energy (\ref{b1}) to the experimental
strength of the logarithmic singularity yields \cite{fl91} $N_0^{1/3}
\approx 50$. The phase $\phi$ should be approximately constant within
$V_0$ (see App.\ \ref{A}). It appears, therefore, tempting (but also
speculative) to set $|\nabla\phi|_{\rm max} \sim V_0^{-1/3}$ and to use
the relation ${\bf u}_{\rm s} = \hbar\, \nabla\phi/m$ for the velocity
of a potential superfluid flow. The maximum (or critical) velocity for
such a flow would then be
\begin{equation}
\label{b4}
v_{\rm crit} = \frac{\hbar \,|\nabla\phi|_{\rm max}}{m}
\sim \frac{\hbar }{m\, V_0^{1/3}} 
= \frac{\hbar }{m\, v^{1/3}}\frac{1}{N_0^{1/3}}
\approx  0.8\,\frac{\rm m}{\rm s} \,.
\end{equation}
In this way the finite size $V_0$ renders a possible connection between
the ``natural'' velocity scale $\hbar /(m \, v^{1/3})$ and realistic
values for the critical velocity. Using Eq.\ (\ref{b3}) we obtain
$v_{\rm crit} \sim |t|^{-2/3}\,{\rm m}/{\rm s}$ for the critical
velocity near the transition point.
\end{enumerate}

\section{Evaluation of $J(\tau)$}
\label{C}
We determine the critical ($\tau\ll 1$) behavior of the integral
\begin{equation}
\label{c1}
J(\tau) = \int_0^\infty \! dx \,x^2\,
\frac{\ln [\exp(x^2 + \tau^2)-1]}{\exp(x^2 + \tau^2)-1} \, .
\end{equation}
Writing $\ln [\exp(x^2 + \tau^2) - 1 ] = \ln [1- \exp(-x^2 - \tau^2)] +
x^2 + \tau^2$ we obtain
\begin{equation}
\label{c2}
J(\tau) = \int_0^\infty \! dx \,
\frac{x^2 \, \ln [1- \exp(-x^2 - \tau^2) ] }{ \exp(x^2 + \tau^2)-1}
+ \int_0^\infty \! dx \;
\frac{x^2 \, (x^2  + \tau^2)  }{ \exp(x^2 + \tau^2)-1} \, .
\end{equation}
The second integral yields a result that is of the form
\begin{equation}
\label{c3}
R_0 = \mbox{const.}+ {\cal O}(\tau^2).
\end{equation}
In the following $R_0$ stands for any expression of this kind. By
${\cal O}(\tau^2)$ we mean terms that are proportional to $\tau^2$ or
to higher powers of $\tau$.

In the first integral in Eq.\ (\ref{c2}) we use the following
expansions into powers of $y=x^2 + \tau^2$ or $\exp(-y)$:
\begin{equation}
\label{c4}
\frac{ \ln [1- \exp(-y) ] }{ \exp(y)-1}
=\left\{ \begin{array}{lc}
\displaystyle
\frac{\ln (y)}{ y} - \frac{ \ln (y) + 1}{2} +
\frac{ y \, (2 \ln (y) +7 )}{24} \pm
\ldots & (y\le 1)
\\[4mm]
\displaystyle
-\exp(-2y) - \frac{3 \exp(-3y)}{2} - \frac{ 11\exp(-4y)}{6} \pm 
\ldots & (y\ge 1) .
\end{array}\right.
\end{equation}
We divide the integration into one part from 0 to 1 and another part
from 1 to $\infty$, and insert the appropriate expansion (\ref{c4}).
All terms except the one with $\ln(y)/y$ yield contributions of the
form (\ref{c3}). We evaluate the remaining integral over $x^2\ln(y)/y =
\ln(y) - \tau^2\ln(y)/y$,
\begin{eqnarray}
J(\tau) &=& \int_0^1 \! dx \, \ln (x^2 + \tau^2)  
- \tau^2 \int_0^1 \! dx\,\frac{ \ln (x^2 + \tau^2)}{x^2 + \tau^2} + R_0
\nonumber\\[1mm]
&=& \pi\tau  - \tau^2 \int_0^\infty \! dx \,
\frac{ \ln (x^2 + \tau^2)  }{x^2 + \tau^2} + R_0 .
\label{c5}
\end{eqnarray}
The first integral yielded $ \pi \tau + R_0$ (see number 2733 in Ref.\
 \cite{gr80}). Because of $\int_1^\infty \!dx \; \ln (x^2 +
\tau^2) /(x^2 + \tau^2) = R_0$ the upper limit of the remaining
integral could be set equal to infinity. In this integral we substitute
$x = \tau z$:
\begin{eqnarray}
J(\tau) &=&
\pi \tau - \tau\int_0^\infty \! dz \, \frac{\ln (\tau^2 )}{1 + z^2}
- \tau \int_0^\infty \! dz \,
\frac{\ln (1+z^2)}{1 + z^2 } + R_0
\nonumber\\[1mm]
&=&   J(0) - \pi \tau  \ln \tau + \pi \tau \big[1- \ln (2)\big]
+ {\cal O}(\tau^2) .
\label{c6}
\end{eqnarray}
The first integral is elementary, the second one may be found under
number 4.295 in Ref.\  \cite{gr80}. A numerical integration yields
$J(0)\approx -1.183$.
\end{appendix}

\end{document}